\documentclass[reprint,superscriptaddress,nofootinbib,twocolumn,amsmath,amssymb,aps,prl]{revtex4-2}

\usepackage{graphicx}% Include figure files
\usepackage{dcolumn}% Align table columns on decimal point
\usepackage{bm}% bold math
\usepackage[colorlinks]{hyperref}

\newcommand{\braket}[1]{\langle#1\rangle}

\newcommand{\ex}[1]{e^{#1}}

\begin{document}

\preprint{APS/123-QED}

\title{Engineering Photon Statistics of Spatial Light Modes}

\author{Mingyuan Hong}
\email{mhong2@lsu.edu}
\affiliation{Quantum Photonics Laboratory, Department of Physics \& Astronomy, Louisiana State University, Baton Rouge, LA 70803, USA}

\author{Ashe Miller}
\affiliation{Quantum Photonics Laboratory, Department of Physics \& Astronomy, Louisiana State University, Baton Rouge, LA 70803, USA}

\author{Roberto de J. Le\'on-Montiel}
\affiliation{Instituto de Ciencias Nucleares, Universidad Nacional Aut\'onoma de M\'exico, Apartado Postal 70-543, 04510 Cd. Mx., M\'exico}

\author{Chenglong You}
\affiliation{Quantum Photonics Laboratory, Department of Physics \& Astronomy, Louisiana State University, Baton Rouge, LA 70803, USA}

\author{Omar S. Maga\~na-Loaiza}
\affiliation{Quantum Photonics Laboratory, Department of Physics \& Astronomy, Louisiana State University, Baton Rouge, LA 70803, USA}

\date{\today}

\begin{abstract}

The nature of light sources is defined by the statistical fluctuations of the electromagnetic field. As such, the photon statistics of light sources are typically associated with distinct emitters. Here, we demonstrate the possibility of producing light beams with various photon statistics through the spatial modulation of coherent light. This is achieved by the sequential encoding of controllable Kolmogorov phase screens in a digital micromirror device. Interestingly, the flexibility of our scheme allows for the arbitrary shaping of spatial light modes with engineered photon statistics at different spatial positions. The performance of our scheme is assessed through the photon-number-resolving characterization of different families of spatial light modes with engineered photon statistics. We believe that the possibility of controlling the photon fluctuations of the light field at arbitrary spatial locations has important implications for quantum spectroscopy, sensing, and imaging. 

\end{abstract}

\maketitle

Manipulation of the spatial properties of light enables the development of novel imaging systems, communication protocols, and schemes for quantum information processing \cite{MaganaLoaiza2019,Willner2015,Flamini2019}. In quantum optics, a photon's transverse spatial degree of freedom  serves as a flexible platform to test complex quantum information protocols in a relatively simple fashion \cite{MaganaLoaiza2019, Willner2015}. These schemes rely on the spatial modulation of photons to produce and characterize diverse families of spatial photonic modes \cite{Arrizon2007, Mirhosseini2013}. Interestingly, spatial light modulators have enabled the engineering of complex spatial photonic wavefunctions through the control of the spatial, temporal, and polarization properties of photons \cite{Zhan2009,Aiello2015,Beckley2010,HashemiRafsanjani2015}. Nevertheless, the possibility of using spatial modulation to manipulate the quantum statistical fluctuations of the electromagnetic field, that define the nature of light sources, remains unexplored \cite{Mandel1979,Gerry2004, MaganaLoaiza2019npj,You2020a, You2021}.

\begin{figure*}[!htbp]
  \centering
  \includegraphics[width=0.85\textwidth]{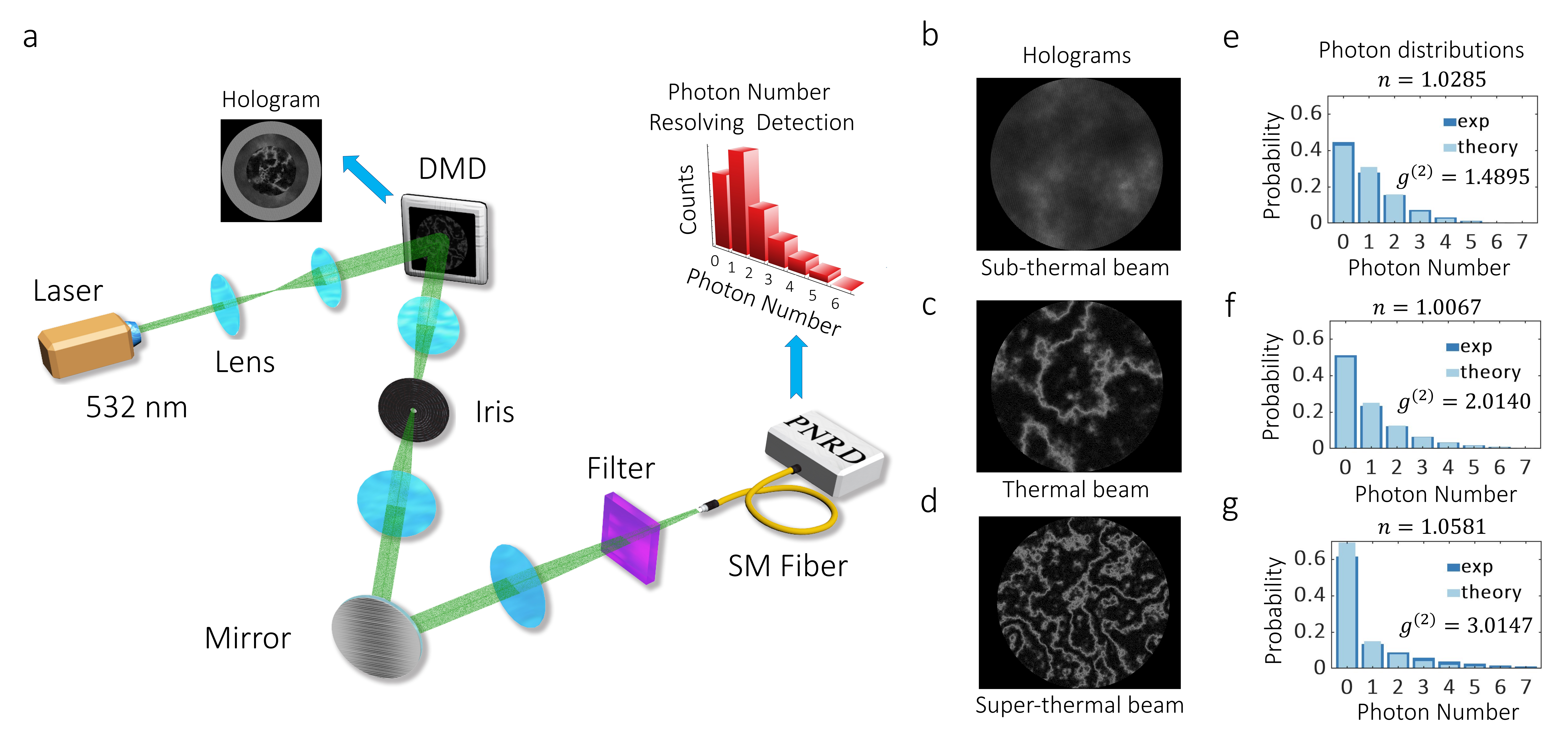}
  \caption{The experimental setup for the engineering of photon statistics of spatial light modes is shown in \textbf{a}. The experiment is performed using a green laser centered at 532 nm. The initial beam is spatially cleaned and expanded using a 4-f system to match the size of the digital micro-mirror device (DMD). Another 4-f system and an iris are used to filter the first diffraction order of the beam reflected off the DMD. The Kolmogorov phase encoded in the filtered light beam induces the engineered photon statistics. We control the mean photon number of the beam using a neutral density filter. Finally, we measure the photon statistics using a single-mode (SM) fiber, an avalanche photodiode (APD) and a time tagger which are used to perform photon-number-resolving detection (PNRD) \cite{You2020a}. The panels from \textbf{b}-\textbf{d} show a sample frame of the holograms displayed on DMD. The Fried's parameter $r_0$ are 12, 0.8 and 0.32 $\mu$m for these holograms, respectively. Panels \textbf{e}-\textbf{g} depict the corresponding photon number statistics. Surprisingly, these engineered light beams can exhibit sub-thermal, thermal, or super-thermal photon statistics. We further quantify these photon fluctuations using the degree of second-order coherence $g^{(2)}$.}
\label{setup}
\end{figure*}

Here, we demonstrate the possibility of using spatial modulation to control the photon fluctuations of the light field, enabling the engineering of spatial light modes with tailored statistics \cite{Arecchi1965,Kondakci2017,anno2006}. Specifically, we prepare Laguerre-Gauss and Hermitte-Gauss modes with sub-thermal, thermal, and super-thermal photon statistics at different pixels. This is achieved by encoding phase screens possessing Kolmogorov statistics in a digital micromirror device (DMD) \cite{Harding1999, Frisch1995}. As such, our scheme enables the engineering of photon statistics for spatial light modes at a kiloHertz rate. In our experiment, we quantify the quality of the produced beams by  photon-number-resolving detection \cite{You2020a, Bhusal2022}. Our results suggest that our modulation scheme might have important implications for other research fields such as quantum spectroscopy, sensing, and imaging \cite{MaganaLoaiza2019,Flamini2019,Lee2021}.

Our technique for engineering the photon statistics of spatial light modes relies on the dynamic modulation of a laser beam's spatial properties of coherence. For this purpose, we use random phase screens generated through the Kolmogorov model of turbulence \cite{Harding1999,Frisch1995}. This model enables us to quantify the spatial coherence properties of the modulated beams. While these random phase screens could be encoded in spatial light modulators (SLMs) \cite{Rodenburg_2014}, we can exploit the speed of DMDs to shape the spatial profile of light at a rate of approximately 12 kHz \cite{Mirhosseini2013}.  In our work, the encoding of Kolmogorov phase screens in a DMD is achieved through amplitude-only spatial light modulation \cite{Mirhosseini2013}. This encoding approach allows one to produce light beams of the form $A(x,y)\ex{i\Phi(x,y)}$, where $A(x,y)$ represents the field amplitude as a function of the transverse spatial coordinates $x$ and $y$, and $\Phi (x,y)$ supplies the corresponding phase information. To achieve simultaneous amplitude and phase modulation, we use two dimensional binary amplitude holograms of the form $T(x,y)=[1+\text{sgn}(\text{cos}[2\pi x/x_0+\pi p(x,y)]-\text{cos}[\pi w(x,y)])]/2$, where sgn() represents the sign function. The position $p(x,y)=\text{arcsin}[A(x,y)]/\pi$  and width $w(x,y)=\Phi(x,y)/\pi$ of the binary grating $T(x,y)$, with period $x_0$, define the corresponding amplitude and phase of the light field shaped by the DMD in its first diffraction order. We define the amplitude $A(x,y)$ and phase $\Phi(x,y)$ through the Kolmogorov model of turbulence \cite{Harding1999,Frisch1995}. Interestingly, the Kolmogorov distortions of the reshaped beams erase the coherence properties of coherent laser fields \cite{Rodenburg2014,Arecchi1965}. The degree of transverse distortion is quantified through the Fried's parameter $r_0$, which is also called Fried's coherent length. Here, we keep  $A(x,y)$ constant and control the amount of randomness in each phase screen through $r_0$. This is achieved by adjusting the distance among the superpixels that define the phase cells as well as their size. Specifically, the Kolmogorov phase screen is given by
\begin{equation}
    \Phi=\text{Re}(\mathcal{F}^{-1}(\mathbb{M}\sqrt{\phi})),
\end{equation}
where $\mathbb{M}$ denotes an encoded random matrix, and the approximated power spectral density $\phi\approx0.023r_0^{-\frac{5}{3}}f^{-\frac{11}{3}}$ \cite{Rodenburg_2014}. In this case, Re() indicates the real part of the complex number, $\mathcal{F}^{-1}$ the inverse Fourier transform, and $f$ the spatial frequency of the light field. Note that the mechanism to generate light with different photon statistics in our setup is similar to that relying on a rotating ground glass \cite{Arecchi1965, Li2020}. More specifically, the total electromagnetic field collected by the detector contains contributions from sub-fields produced by multiple emitters at random spatial positions or pixels. According to the central limit theorem, the detected photon number $n$, obeys a negative-exponential distribution similar to the thermal distribution \cite{Mandel1959}. Based on this principle, by carefully adjusting the strength of random phases, we can exert control of the photon fluctuations of the electromagnetic field at arbitrary spatial locations. 

We first explore the modification of photon statistics using the experimental setup depicted in Fig. \ref{setup}a. For this purpose, we display videos on the DMD (DLP LightCrafter 6500 Evaluation Module) at a rate of 60 frames per second. Each frame of the video contains random transverse structures characterized by the Kolmogorov model \cite{Harding1999}. These videos are made up with frames containing values of either 0 and 255 (in uint8 format) in each pixel. Based on the DMD's working principle, only pixels with 255 are able to reflect photons to the detector. This setup allows us to encode random Kolmogorov phases with different fluctuation strengths. We show sample holograms in Fig. \ref{setup}b-d. The encoded phase distributions induce photon fluctuations upon propagation \cite{MaganaLoaiza2022}. This mechanism allows us to use the encoded holograms to engineer spatial modes with sub-thermal, thermal, or super-thermal statistics. The photon statistics of the engineered modes are quantified through the degree of second-order coherence $g^{(2)}=1+(\braket{(\Delta\hat{n})^2}-\braket{\hat{n}})/\braket{\hat{n}}^2$. It is well-known that coherent light beams are characterized by $g^{(2)}=1$, whereas thermal fields exhibit a $g^{(2)}=2$ \cite{Gerry2004}. In addition, light beams with $1<g^{(2)}<2$ are known as partially coherent or sub-thermal optical fields. Also, optical beams with $g^{(2)}>2$ are called super-thermal light fields \cite{Allevi2017, Lettau2018}. As demonstrated in Fig. \ref{setup}e-g, these random phase screens thermalize the initial coherent laser beam to produce a degree of second-order coherence $g^{(2)}>1$. In our experiment, we are able to preserve the coherent statistics of the initial beam by encoding phase screens with Fried's parameters $r_0$ that tend to infinity \cite{Harding1999}. Furthermore, we thermalize the optical field by reducing $r_0$. This approach increases the value of $g^{(2)}$. As shown in Fig. \ref{setup}e, we generate sub-thermal fields by using phase screens with $r_0=12$ $\mu$m. Moreover, we produce a thermal field characterized by $g^{(2)}=2$ through the encoding of Kolmogorov phase screens with $r_0=0.8$ $\mu$m. Finally, we generate light with super-thermal statistics, this is reported in Fig. \ref{setup}g. As discussed below, we sequentially display empty and random phase screens on the DMD.

\begin{figure}[t!]
  \centering
  \includegraphics[width=0.85\linewidth]{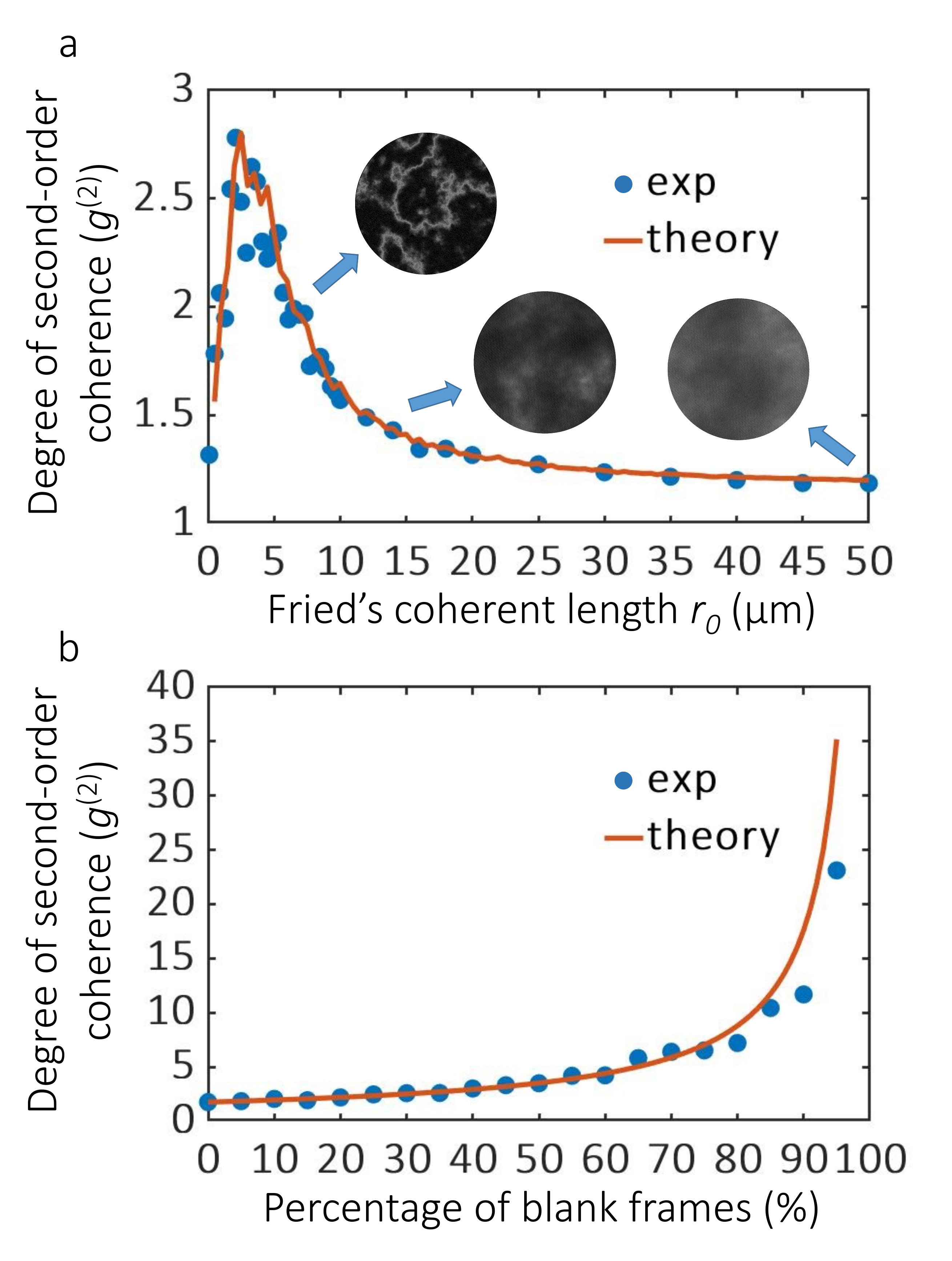}
  \caption{Relation between the second-order degree of coherence $g^{(2)}$ and the encoded Kolmogorov phase screens. Panel \textbf{a} reports $g^{(2)}$ as a function of the Fried's parameter $r_0$. The insets show characteristic holograms corresponding to three Fried's coherence lenghts. The theoretical prediction relies on a Monte Carlo simulation discussed below.
  In panel \textbf{b}, we plot the experimental and theoretical $g^{(2)}$ as a function of the percentage of empty frames. In this experiment, we fix the Fried's parameter $r_0$ to $0.32$ $\mu$m. This theoretical prediction is based on the summation of the original photon statistics and vacuum events with corresponding proportions.}
\label{curve}
\end{figure}

% \begin{figure}[t!]
%   \centering
%   \includegraphics[width=0.85\linewidth]{fig2v1.pdf}
%   \caption{[\textbf{a} in edition]. In panel \textbf{b}, we plot experimental data together with our theoretical prediction for the relation between the degree of second-order coherence ($g^{(2)}$) and the percentage of empty frames in the videos that produce super-thermal statistics photons. In this experiment, the Fried's parameter of the nonempty frames is fixed at 0.32 $\mu$m as introduced in Fig. \ref{setup}d. The theoretical model is based on the photon-number distribution prediction of thermal states combined with vacuum states.}
% \label{curve}
% \end{figure}

We now explore the properties of the produced photon fluctuations as a function of the Kolmogorov phase screens. For this purpose, we display videos containing various phase screens on the DMD and measure the corresponding photon-number distribution in its far-field. These measurements allow us to calculate the $g^{(2)}$ of the shaped beams. In Fig. \ref{curve}a, we report the $g^{(2)}$ dependency on the Fried’s parameter $r_0$. We note that in both limits, specifically, when $r_0$ is tending to 0 or infinity, $g^{(2)}$ is close to 1. This phenomenon implies that in these limits, the photons we measure fluctuate in a similar fashion to those prepared in a coherent state \cite{Gerry2004}. Given the resolution of DMD, 1920 by 1080 pixels, when $r_0$ goes to 0 or infinity, the DMD behaves like a mirror and the statistics of the laser beam are preserved. Interestingly, when $r_0$ increases, $g^{(2)}$ first increases up to 3 and gradually drops back to 1. We theoretically modeled the observed behavior of the $g^{(2)}$ using a Monte Carlo simulation of the experiment \cite{Voelz2011}. Remarkably, this model is in excellent agreement with our experiment. Given the fact that the probability of measuring a vacuum state is larger for super-thermal light than for thermal or sub-thermal light beams \cite{MaganaLoaiza2019npj, Gerry2004}, we study the role of displaying empty frames on the DMD during the sequential encoding of Kolmogorov phase screens. As reported in Fig. \ref{curve}b, the magnitude of the $g^{(2)}$ increases with the number of empty frames. Similarly, we theoretically modeled the behavior of the second-order coherence by increasing the probability of measuring vacuum events. In both panels, the remarkable agreement between our theoretical and experimental results validates the modification of the quantum statistical fluctuations of the light field through its spatial modulation.

The Monte-Carlo simulation was performed by encoding a phase screen with Kolmogorov statistics onto a Gaussian beam. This process allows us to emulate the DMD transformation. We propagate the beam through free space before measuring the mean photon number in the region of the field. This mean photon number is then used to generate the photon number distributions. As such, our scheme shares similarities with the rotating ground glass technique to produce pseudothermal light \cite{Li2020}. This procedure is repeated many times for each $r_0$ value allowing us to calculate the average $g^{(2)}$. This whole process was repeated for a range of $r_0$'s creating the theoretical curve in Fig. \ref{curve}a.

\begin{figure*}[!htbp]
  \centering
  \includegraphics[width=0.85\textwidth]{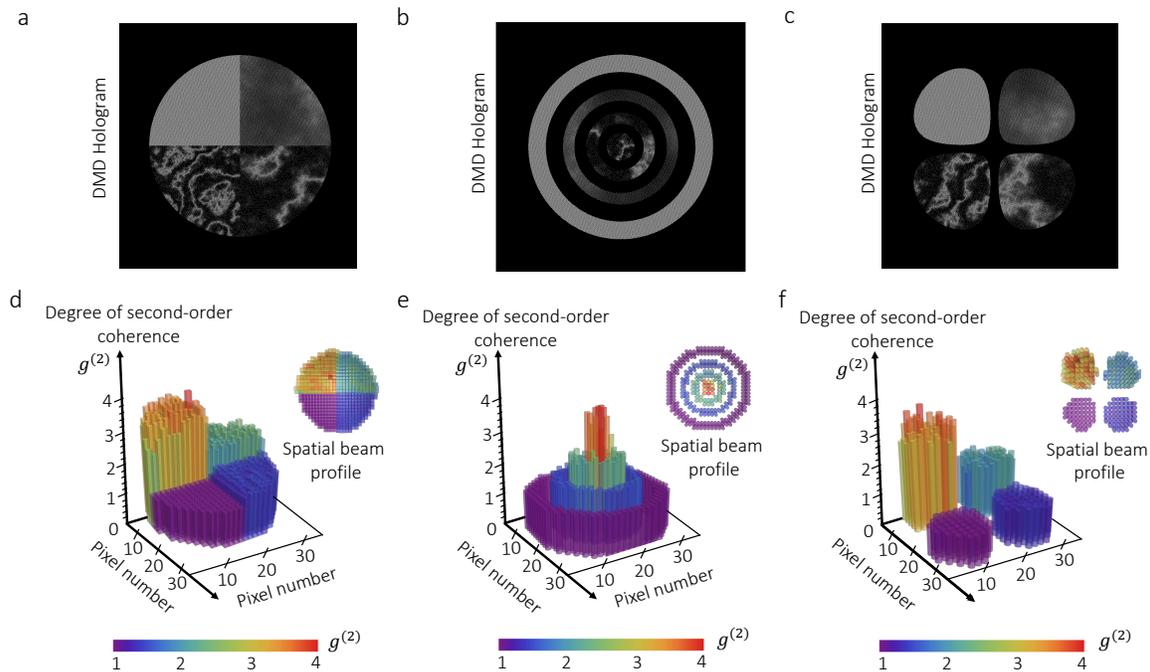}
  \caption{Generation of spatial light modes with arbitrary photon statistics. The panels from \textbf{a}-\textbf{c}
  show three holograms that enable the engineering of optical modes with coherent, partially coherent, thermal, and super-thermal photon statistics. The hologram in \textbf{a} shows a Gaussian mode with four quandrants characterized by different Fried's parameters. Similarly, the sample hologram in \textbf{b} enables one to encode a Laguerre-Gaussian mode LG$_{30}$ with rings characterized by different photon statistics. In \textbf{c}, we show a sample  hologram that enables the engineering of a Hermite-Gaussian TEM$_{11}$ mode with different photon fluctuations. These beams were characterized  by implementing 32$\times$32 raster scanning using a second DMD \cite{You2020a, Bhusal2022}. These measurements allowed us to perform the characterizations reported from \textbf{d} to \textbf{f}. The integration time for each pixel was 3 seconds. These results demonstrate the possibility of using a DMD for  spatial modulation to generate light modes with tailored photon statistics.}
\label{raster}
\end{figure*}

% \begin{figure*}[!htbp]
%   \centering
%   \includegraphics[width=0.85\textwidth]{Fig3t.pdf}
%   \caption{The three kinds of combination of phases producing coherent, sub-thermal, thermal and super-thermal photon statistics are shown in \textbf{a}-\textbf{c}. The four types of grated phases are separated into four sectors in \textbf{a}; four rings with the same size in \textbf{b}; and four parts of Hermite-Gaussian TEM$_{11}$ mode shape in \textbf{c}. To produce coherent states, we set the Fried's parameter to be infinity, while the other three $r_0$'s keep the same as those in Fig. \ref{setup}. By implementing 32$\times$32 raster scanning with a second DMD, we are thus able to get the experimental results depicted in \textbf{d}-\textbf{f}. The measurement for each pixel lasts 3 seconds while our time resolution is 80 picoseconds. By plotting the degree of second-order coherence of every pixel that represents the photon-number distribution, we can clearly reproduce the original shape separation of the \textbf{a}-\textbf{c} holograms.}
% \label{raster}
% \end{figure*}

We now demonstrate the flexibility of our scheme to produce arbitrary spatial modes of light with engineered photon statistics. For this purpose, we divide the DMD screen into different regions. We then display random phase screens characterized by different magnitudes of the Fried's parameter in each region.  In our demonstration, we generated the three spatial modes reported in Fig. \ref{raster}a-c. To measure the photon number distribution of each spatial mode, we image the generated beam onto another DMD using a 4-f system \cite{Bhusal2022}. The second DMD is used to perform raster scanning allowing us to measure the $g^{(2)}$ of the generated beams at each pixel \cite{Bhusal2022}. The corresponding results are presented in Fig. \ref{raster}d-f. These results demonstrate the possibility of engineering the photon statistics and the $g^{(2)}$  for each pixel of a spatial mode. Interestingly, the images of the characterized beams resemble the original spatial distributions of the phase screens in Fig. \ref{raster}a-c. These results validate our method to generate beams of light with tailored photon statistics at arbitrary spatial positions. 

% 
% Specifically, the hologram is separated into four quadrants with equal radians and radii in Fig. \ref{raster}a, to make sure each quadrant receives approximately same amount of photons. Similarly, Fig. \ref{raster}b consists of four rings with the same area, and Fig. \ref{raster}c displays the four paddles of Hermite-Gaussian TEM$_{11}$ mode.
% with $32\times32$ super-pixels. We note these super-pixels is artificially generated by combining 33$\times$33 pixels of the screen. By using raster scanning,

In conclusion, the photon statistics are fundamental quantum mechanical properties of light sources \cite{Arecchi1965,Kondakci2017,anno2006,You2021, Bhusal2022}. Here, we experimentally demonstrated the arbitrary shaping of spatial light modes with engineered photon statistics at different spatial positions. Our work relies on the sequential modulation of coherent laser beams through the encoding of Kolmogorov phase screens in a DMD. As such, our scheme provides a flexible platform for the generation of complex optical beams with multiple degrees of freedom. These optical beams were characterized through photon-number-resolving
detection \cite{You2020a, Bhusal2022}. We believe that the possibility of controlling the photon statistics, at arbitrary spatial locations, has important implications for quantum spectroscopy, sensing, and imaging \cite{anno2006, obrien2009, MaganaLoaiza2019, Degen2017}.

% The initial beam was enlarged with a 4-f system consisting of two lenses, so that the size of the output collimated beam is larger than the digital micro-mirrors device (DMD) screen. Certain videos produced with Matlab were then displayed on the DMD screen which was controlled by a computer software (DLP LightCrafter). The videos were made up with bmp frames (60 fps) containing only 0 and 255 (in uint8 format) in each pixel. Pixels with 0 wouldn’t be activated, while only pixels with 255 were able to reflect photons as those corresponding micro-mirrors would deflect to the angle ($\pm12^{\circ}$) we want. The reshaped beam was then propagated into another 4-f system consisting of two lenses with the same focal length. At the focus point between these two lenses, an iris was mounted at a specific position in order to cut off most of the beam except for the first order which entirely showed the properties of the phase provided by the videos without the disturbance of background coherent light from the laser. The output beam carrying the phase information would then be coupled into a single-mode fiber through another coupling lens and certain ND filters to reduce the mean photon number. Proper amount of photons were then gathered by an avalanche photodiode (APD) and analyzed by a time tagger to give us the quantum statistics of photon number distributions with the method of photon number resolving detection (PNRD). The distributions shown in Fig. \ref{setup}e-g and Fig. \ref{raster} demonstrated our successful modulation of photon statistics of spatial light.

We acknowledge funding from the U.S. Department of Energy, Office of Basic Energy Sciences, Division of Materials Sciences and Engineering under Award DE-SC0021069. 

%\section{COMPETING INTERESTS}
%The authors declare no competing interests.

%\section{DATA AVAILABILITY}
%The data sets generated and/or analyzed during this study are available from the corresponding author or last author on reasonable request.

%\section{CODE AVAILABILITY}
%The code used to analyze the data and the related simulation files are available from GitHub repository at ...

\bibliography{main}% Produces the bibliography via BibTeX.
\end{document}